\documentclass[a4paper,11pt]{article}
\pdfoutput=1 
\usepackage{jheppub} 
\usepackage[T1]{fontenc} 
\usepackage{indentfirst}
\usepackage{comment}
\usepackage{pdfpages}
\usepackage{xcolor}

\usepackage{feynmp}
\title{\boldmath Higgs boson decay into gluons in a 4D regularization: IR cancellation without evanescent fields to NLO}

\author[a]{Ana Pereira,}%\note{}}
\affiliation[a]{~CFisUC, Department of Physics, University of Coimbra, 3004-516 Coimbra,  Portugal}

\author[b]{Adriano Cherchiglia,}%\note{}}
\affiliation[b]{~Instituto de Física Gleb Wataghin -- Universidade Estadual de Campinas, Campinas-SP, Brasil}

\author[c]{Marcos Sampaio,}%\note{}}
\affiliation[c]{Universidade Federal do ABC, 09210-580 , Santo André, Brasil}

\author[a]{Brigitte Hiller}%\note{}}
\abstract{Higgs decay using an effective Higgs--Yang-Mills interaction in terms of a dimension five operator as well as usual QCD interactions is revisited in the context of Implicit Regularization (IReg) and compared with conventional dimensional regularization (CDR) and four dimensional helicity (FDH) schemes.
The decay rate for Higgs$\rightarrow 2$ gluons  is calculated in this strictly four-dimensional set-up to  $\alpha_s^3$ order in the strong coupling. Moreover we include joint processes that contribute at the same perturbative order in the real emission channels consisting of 3 gluons as well as gluon quark-antiquark final states with light (zero mass) quarks.
Unambiguous identification and separation of UV from IR divergences is achieved putting at work the renormalization group scale relation inherent to the method. UV singularities are removed as usual by renormalization, the IR divergences are cancelled due to the method’s compliance with the Kinoshita-Lee-Nauenberg (KLN) theorem. The remaining finite integral contributions are thus rendered more amenable for numerical evaluation. Most importantly, we verify that no evanescent fields such as $\epsilon$-scalars need be introduced as required by some mixed regularizations that operate partially in the physical dimension.}

\begin{document} 
\maketitle
\flushbottom

\section{Introduction} \label{Introduction}

The Standard Model of Particle Physics (SM) is a good description of the physics
at and below the electroweak scale, the discovery of the Higgs boson at the Large Hadron Collider (LHC) being the confirmation of this framework \citep{ATLAS:2012yve}. It is also clear that the SM does not provide
a complete description of particle interactions. Phenomena such as dark matter and dark energy, a consistent quantum theory of gravity,  the stability of electroweak vacuum  up to the Planck scale and the hierarchy problem, to name a few, motivated the development of models beyond
the SM (BSM).

The Future Circular Collider aims at reaching  collision energies of around 100 TeV demanding  higher precision in theoretical computations for Standard
Model phenomena and electroweak pseudo-observables (EWPOs). 
New calculational methods have been developed through computer algebraic algorithms for analytical and numerical
methods. At a given loop order, the complexity is measured  by the number of virtual massive particles in the evaluation of
Feynman integrals. On the other hand, a theoretical library for perturbation theory calculations of Feynman integrals in closed form beyond
one loop does not exist \citep{2021arXiv210611802H}. While numerical integration methods seems to be the main tool to address those problems, analytical techniques play  important role to systematically and unambiguously remove infrared and ultraviolet divergences from physical observables.

Regularization frameworks that operates partially or entirely in the physical dimensional can bring some advantages and simplifications in the evaluation of Feynman amplitudes  and extensions such as dimensional reduction (DRED), Four Dimensional Helicity scheme (FDH), Implicit Regularization
(IReg), among others \citep{2017EPJC...77..471G} have been constructed. 
\begin{itemize}
\item CDR (Conventional Dimensional Regularization): internal and external gluons are all treated as d-dimensional.
\item HV ('t Hooft Veltman scheme): internal gluons are d-dimensional and external ones are strictly 4-dimensional.
\item DRED (original Dimensional Reduction”): internal and external gluons are all treated as quasi-4-dimensional.
\item FDH (Four-Dimensional Helicity): internal gluons are treated as quasi 4-dimensional and external ones are treated as strictly 4-dimensional.
\item IReg (Implicit Regularization): all fields as well as internal momenta are defined in the physical dimension.
\end{itemize}

Therefore IReg acts directly on the dimension of the theory and can be systematically implemented to all orders in perturbation theory \cite{2008EPJC...55..667D,Cherchiglia:2010yd}. 

CDR relies on the assumption the
the quantities of interest depend smoothly on the spacetime dimensionality. Such an assumption leads to ambiguities in chiral and supersymmetric quantum field theories. For instance, the
expression for chiral anomalies are specific for a given dimension cannot
be unambiguously analytically continued in dimensionality. Mixed schemes such as DRED and FDH aim to extend the scope of CDR where it potentially fails.

%%%%%%
In mixed regularization schemes that operate partially in the physical dimension such DRED or FDH, an auxiliary space which has the characteristics of a $4$-dimensional space, $Q4S$, is introduced. Such quasi-$4$-dimensional space is decomposed
as $Q4S=QdS\oplus Q n_{\epsilon} S$, where $QdS$  is formally $d$-dimensional and its complement $Q n_{\epsilon} S$ has dimension $n_{\epsilon}=4-d$
\cite{2014PhLB..733..296G}. The metric tensor for the original 4-dimensional space 4S is denoted by $\Bar{g}^\mu_\nu $ whereas the metric tensor of the spaces $Q4S$, $QdS$ and
$Q n_{\epsilon} S$ are respectively written as $g^\mu_\nu$, $\hat{g}^\mu_\nu$ and
$\tilde{g}^\mu_\nu$.  They satisfy
\begin{equation}
g^\mu_\nu =\hat{g}^\mu_\nu+\tilde{g}^\mu_\nu, \,\,
g^{\mu\nu}\tilde{g}^\rho_\nu =\tilde{g}^{\mu\rho}, \,\,
g^{\mu\nu}\hat{g}^\rho_\nu =\hat{g}^{\mu\rho}, \,\,
\hat{g}^{\mu \rho}\Bar{g}_\rho^\nu = \Bar{g}^{\mu \nu}, \,\, 
\hat{g}^{\mu\nu}\tilde{g}^\rho_\nu = 0 , 
\end{equation}
with
\begin{equation}g^\mu_\mu=4, \, \tilde{g}^\mu_\mu=n_{\epsilon}=2\epsilon \,\,\, \text{and} \,\,\, \hat{g}^\mu_\mu=d.
\end{equation}
Furthermore, mathematical consistency and d-dimensional gauge invariance require that $Q4S \supset
QDS \supset 4S$ and forbid to identify $g^{\mu \nu}$ as $\Bar{g}
^{\mu \nu}$ \citep{2006PhLB..634...63H}.

Due to the
decomposition of $Q4S$, the gauge field is split as 
$A^a_\mu = \hat{A} ^a _\mu + \epsilon ^a _\mu$, where
$\hat{A}^a_\mu \in QdS$ and $\epsilon ^a_\mu \in Q n_{\epsilon} S$. The
$\epsilon$-dimensional field $\epsilon ^a_\mu$ is a scalar under
$d$-dimensional Lorentz transformations and transforms in the adjoint representation.  Due to this decomposition, the Lagrangian is modified to incorporate $\epsilon$-scalars, giving rise to extra Feynman diagrams with
$\epsilon$-scalars, which contribute additional finite terms to
divergent loop amplitudes \cite{2015EPJC...75..424B} 
\footnote{Interestingly in \cite{2018JHEP...08..026S},  $\epsilon$-scalars are somewhat integrated out in
the language of effective field theories. In the limit $\epsilon \rightarrow 0$ this effectively results in a change of the regularization scheme from DRED to CDR}.  

Ideally a fully mathematical consistent regularization scheme that prevents the emergence of symmetry breaking terms or spurious anomalies and that is valid to arbitrary higher order is necessary. IReg is advantageous in the sense that gauge symmetry breaking terms are linked to momentum routing violation in Feynman diagram loops. Such symmetry breaking terms accompany surface terms whose structure is built to arbitrary loop order, in a regularization independent fashion. Ultraviolet renormalization is constructed by subtracting loop integrals which need not be explicitly evaluated to define renormalization group functions. On the other hand an invariant regularization scheme must comply with infrared finiteness as stated by the Kinoshita–Lee–Nauenberg theorem \citep{PhysRev.133.B1549}, \citep{osti_4784262}. KLN theorem, in a nutshell, states that in gauge theories the infrared divergences coming from loop integrals are canceled by IR divergences coming from phase space integrals.
As a result it is of theoretical interest to test the applicability of IReg in a practical calculation involving IR divergences that only cancel at the level of cross sections or decay rates and verify the KLN theorem.

The $H \rightarrow gg$ decay described by an effective model in which the top quark is integrated out, provides a simple and reliable model to test this regularization scheme, as suggested in \citep{Broggio2015ComputationO}. Such a model  has been used in \citep{2014EPJC...74.2686P} to test the four-dimensional
regularization/renormalization (FDR) method.

The main goal of this work is the computation of the total decay rate of this process in IReg to NLO to show that, contrarily to regularization schemes that work partially in the physical dimension, no modifications to the Lagrangian such as $\epsilon$-fields are required. This amounts to a significant simplification which will have a great effect in calculations beyond NLO. We will compute the one-loop virtual diagrams that arise from this effective model and use IReg to extract the UV divergences which are absorbed in the process of renormalization, and use the remaining UV finite amplitudes to obtain the regularized virtual decay rate. Then we apply the spinor-helicity formalism to compute the real diagrams of the process and obtain the real decay rate. It is expected to have a cancellation of the IR divergences when combining these decay rates and the final result must be finite.

Our work is organized as follows: in section \ref{The rules of IReg} we briefly review the IReg method, which is applied in section \ref{Higgs decays} to the decay $H \longrightarrow gg$ at NLO. Section \ref{sec:comparison} is devoted to a comparison of our results to dimensional schemes. Finally, we conclude in section \ref{sec:conclusion}.

\section{The IReg method: UV/IR identification and UV renormalization}
\label{The rules of IReg}

IReg is a regularization method that operates on the momentum space and was shown to respect unitarity, locality and Lorentz invariance \citep{2011IJMPA..26.2591C}. This procedure operates on the specific physical dimension of the theory, therefore we do not need to extend the space-time dimensions. IReg also does not require any changes to the Lagrangian and can be applicable to arbitrary $n$-loop calculations, making it an alternative to dimensional schemes. In this work we will be concerned with one-loop examples only, a recent review of the method applicable to $n$-loop amplitudes can be found in
\cite{2021Symm...13..956A}.
%\cite{Arias-Perdomo:2021inz}

The main idea of IReg is to use an algebraic identity at integrand level recursively until the UV divergent behavior is only present in irreducible loop integrals that depend on internal momentum. In this way, the UV finite content of the amplitude (that may still be IR divergent) will contain denominators with dependence on physical parameters (external momenta and masses). To be concrete, we illustrate the method with the toy integral
\begin{equation}
   \int_k \dfrac{1}{k^2 (k-p)^2}, \quad \int_k \frac{dk^{4}}{(2\pi)^{4}}.
\end{equation}

We start by introducing a infrared regulator $\mu$ in the denominator like
\begin{equation}
     \lim_{\mu\rightarrow 0}\int_k \dfrac{1}{(k^2-\mu^2) [(k-p)^2 - \mu^2]}.
\end{equation}
In the case of IR safe integrals, the regulator $\mu$ is needed to avoid spurious IR divergences in the course of the evaluation. It will cancel in the end result. In the case of IR divergent integrals, the $\mu$ will survive and parameterize the IR divergences.

By power counting, we notice that as $k \longrightarrow \infty$ this integral diverges, but there is a dependence both on internal and external momenta. We want to isolate the UV divergent content in an integral that is solely dependent on the internal momentum \textit{k}. We notice that it is possible to rewrite the portion of the integrand that depends on external momentum as
\begin{equation} \label{2.8}
   \dfrac{1}{(k-p)^2 - \mu^2} = \dfrac{1}{k^2-\mu^2} + \dfrac{2k\cdot p - p^2}{(k^2-\mu^2)((k-p)^2-\mu^2)},
\end{equation}
where in the second term we diminish one order of divergence and in the first one we have an integral depending only on the internal momentum as we wanted. As a result, this identity can be used to manipulate the expression and isolate the UV divergent content in integrals depending only on the internal momentum. The procedure exemplified above works in general, by repetitive usage of equation \ref{2.8}.

Following the separation of the divergences of the amplitude, the UV divergent content of the amplitude can be expressed as integrals whose denominator is only dependent on the internal momentum \textit{k}. These integrals are classified as Basic Divergent Integrals (BDI's) and they can take either logarithmic or quadratic forms which are respectively

\begin{equation}
    I_{log}^{\nu_1...\nu_{2r}}(\mu^2) = \int_k \dfrac{k^{\nu_1}...k^{\nu_{2r}}}{(k^2 - \mu^2)^{r+2}},\quad\mbox{and}\quad I_{quad}^{\nu_1...\nu_{2r}}(\mu^2) = \int_k \dfrac{k^{\nu_1}...k^{2r}}{(k^2 - \mu^2)^{r+1}}.
\end{equation}

Any BDI with odd power of \textit{k} in the numerator is automatically zero once the integral goes over the entire space-time and all the denominators have even powers of \textit{k}. These BDI's are written in terms of Lorenz indices and can be rewritten as scalar integrals, multiplying metric tensors, after setting Surface Terms (ST) to zero, as we will explain shortly.
Scalar logarithmic and quadratic divergent integrals are finally given as

\begin{equation} 
    I_{log}(\mu^2) = \int_k \dfrac{1}{(k^2 - \mu^2)^2},\quad\mbox{and}\quad I_{quad}(\mu^2) = \int_k \dfrac{1}{(k^2 - \mu^2)}.
\end{equation}

%\subsection{Surface terms (ST's)}

Previous work has shown that IReg preserves the symmetries of the system, such as Lorenz invariance and abelian gauge invariance to arbitrary loop order \cite{2012PhRvD..86b5016F,2016PhRvD..94f5023V,Cherchiglia:2010yd}. In the method, symmetry breaking terms can be expressed as a well defined difference between divergent integrals with the same superficial degree of freedom.
These are called surface terms and they are not originally fixed, which indicates that they are related to momentum routing invariance in Feynman diagrams (the possibility to perform a shift in the integration variables). As their value is associated with symmetry breaking terms, they play a critical role in IReg for the preservation of the symmetries of the system and we must carefully choose a value that allows the symmetries of the underlying theory to be preserved.
Nonetheless, in a constrained version of IReg, it has been proven that these regularization dependent surface terms may be set to zero, complying with gauge invariance, \citep{2018PhRvD..98b5018B,2021EPJC...81..468C}. This will actually allow us to reduce BDI's with Lorenz indices $\nu_1...\nu_{2r}$ to linear combinations of scalar integrals with the same degree of divergence divergence (multiplied by metric tensor combinations), plus well defined surface terms (ST's).
Generally in the four dimensional Minkowskian space-time a surface term of order \textit{j} can be written as

\begin{equation}
    \Gamma_i^{\nu_1 ... \nu_j} = \int_k \dfrac{\partial}{\partial k_{\nu_1}} \dfrac{k^{\nu_2}...k^{\nu_j}}{(k^2 - \mu^2)^{(2+j-1)/2}},
\end{equation}
with \textit{k} being the internal momentum and $\mu$ the infrared regulator introduced earlier. The general formula allows for the computation of any order surface terms. For instance, one obtains

\begin{equation}
    \Gamma_0^{\mu \nu} = \int_k \dfrac{\partial}{\partial_\mu} \dfrac{k^\nu}{(k^2 - \mu^2)^2} = 4 \Big( \dfrac{g_{\mu \nu}}{4} I_{log}(\mu^2) - I_{log}^{\mu \nu}(\mu^2) \Big) = 0
\end{equation}

%\subsection{Renormalization scale}
Finally, when considering IR-safe integrals, one must still take the limit in which the $\mu$ infrared regulator is set to zero. In this case, one rewrites the BDI's in terms of a positive arbitrary constant $\lambda$ which will play the role of the renormalization group scale. It is achieved by using the equation below

\begin{equation} \label{2.19}
    I_{log}(\mu^2) = I_{log}(\lambda^2) + b \ln \Big( \dfrac{\lambda^2}{\mu^2} \Big), \quad b = \dfrac{i}{(4 \pi)^2}.
\end{equation}
%where the constant is $b = \dfrac{i}{(4 \pi)^2}$. 
It is worth noting that a minimal subtraction renormalization scheme emerges naturally from this formalism, in which the infinite divergences that depend only on the internal momentum are subtracted from the theory.
This means that the $I_{log}(\lambda^2)$ will be subtracted via renormalization whereas the IR divergent part $\ln(\mu^2)$ will cancel in the final amplitude for infrared safe processes and in the cross section/decay rate otherwise, which are IR-safe observables.

\section{NLO corrections to \texorpdfstring{$H\rightarrow gg$}{TEXT} in the large top mass limit}
\label{Higgs decays}

As discussed in the introduction, we will use the  decay $H\rightarrow gg$ as working example to test the method of IReg in the presence of both UV/IR divergences using an effective non-abelian field theory approach. Our objective is twofold: (a) the renormalization of an effective field theory is highly non-trivial, in particular, when considering alternatives to CDR \cite{Broggio2015ComputationO, 2014PhLB..733..296G}. Thus, it is essential to understand how IReg can be applied in this context. (b) The presence of both UV/IR divergences requires a precise match between virtual and real contributions in order to find a finite and {\textbf{regularization independent}} result. Thus, it is a stringent test for any regularization scheme. Finally, the decay $H\rightarrow gg$ has served as a benchmark for other regularization methods, allowing for a clear comparison among methods\cite{Broggio2015ComputationO, 2014PhLB..733..296G,2014EPJC...74.2686P}. 

As usual, since the decay we consider is mainly due to the top quark loop, it is reliable to consider the limit in which its mass is infinite. Thus, we have to add the following term to the massless QCD Lagrangian \cite{1997PhRvD..55.4005K,Schmidt1997HggggqqbarAT}
\begin{equation}
    L_{eff} = -\dfrac{1}{4}AH G_{\mu \nu}^{a} G^{a,\mu \nu},
\end{equation}
where \textit{H} represents the Higgs boson field, $G_{\mu \nu}^a$ is the field strength of the $SU(3)$ gluon field given by
\begin{equation}
    G_{\mu \nu}^a = \partial_\mu A_\nu^a - \partial_\nu A_\mu^a + g_{s} f^{abc} A_\mu^b A_\nu^c
\end{equation}
and $f^{abc}$ are the anti-symmetric $SU(3)$ structure constants. The effective coupling \textit{A} can be obtained by performing the matching of the full theory to its effective version, being given by \cite{1995NuPhB.453...17S,DAWSON1991283}
\begin{equation} \label{eq4.11}
    A = \dfrac{\alpha_s}{3 \pi v}\Big( 1 + \dfrac{11}{4} \dfrac{\alpha_s}{\pi}\Big),
\end{equation}
where \textit{v} is the electroweak vacuum expectation value, $v^2=(G_f\sqrt{2})^{-1}$. The Feynman rules can be straightforward obtained, which we collect in Appendix \ref{Feyn_rules}. For the diagrams involving only gluons the Feynman rules are given by the Yang Mills Lagrangian.

Once the model is defined, we present in the next subsections its UV renormalization at one-loop level, as well as the calculation of the virtual and real contributions for the decay $H\rightarrow gg$.

\subsection{UV renormalization}
As usual we adopt multiplicative renormalization, rewriting the effective Lagrangian as 
\begin{equation} 
    (L_{eff})_{ren} = -\dfrac{1}{4} Z_{\alpha_s} Z_A AH G_{\mu \nu} G^{\mu \nu},
\end{equation}
where $Z_{A}$ and $Z_{\alpha_s}$ are the renormalization constants for the gluon-field and coupling constant respectively. Notice that we do not renormalize the Higgs field, since it can only appear as an external leg in the process we consider. The part of the Lagrangian corresponding to massless QCD is renormalized in the standard way, implying that $Z_{A}$ and $Z_{\alpha_s}$ are already known. In the framework of IReg, they are given in \cite{2006IJTP...45..436S}. 

Thus, at first order on $\alpha_{s}$, $Z_{A}$ and $Z_{\alpha_s}$ are given by

\begin{align}
    Z_A &= 1 + \alpha_s \dfrac{1}{(4\pi)} \dfrac{1}{b} I_{log}(\mu^2)\Big[ \Big( \dfrac{13}{6} - \dfrac{\zeta}{2} \Big)C_A - \dfrac{4}{3} T_F N_F \Big] + O(\alpha_s^2)\\
    Z_{\alpha_{s}} &= 1 - \alpha_s \dfrac{1}{(4\pi)} \dfrac{1}{b} I_{log}(\lambda^2) \Big[ \dfrac{11}{6} C_A - \dfrac{2}{3} T_F N_F \Big] +O(\alpha_s^2)
\end{align}
where $\zeta$ is the gauge parameter, $N_F$ is the number of light quarks flavours and $T_F$ is the trace over colour matrices. Notice that, while for $Z_{\alpha_{s}}$ we have the UV divergence expressed as $I_{log}(\lambda^2)$ as usual (minimal subtraction scheme), for $Z_{A}$ we have $I_{log}(\mu^2)$. This happens because we are considering on-shell gluons, $\mu^{2}$ playing the role of their fictitious mass. % In the case of DReg, this case would correspond to a scaleless integral, which is null in that method. 

Finally, by considering the terms just discussed the counterterm to be added to our process will amount to (we adopt Feynman Gauge, $\zeta=1$)  
\begin{equation}\label{count}
    V_{count} =  \dfrac{\alpha_s}{b \pi} \left[C_A \Big(\dfrac{5}{12} I_{log}(\mu^2) - \dfrac{11}{12} I_{log}(\lambda^2) \Big) - \dfrac{1}{3} T_F N_F \Big( I_{log}(\lambda^2) - I_{log}(\mu^2) \Big) \right] V_0.
\end{equation}
where $V_{0}$ corresponds to the tree-level amplitude for $H\rightarrow gg$.

\subsection{Virtual contributions to \texorpdfstring{$H\rightarrow gg$}{TEXT}} \label{Virtual Decay Rate}

\begin{figure}
    \centering
    \includegraphics[width=\textwidth]{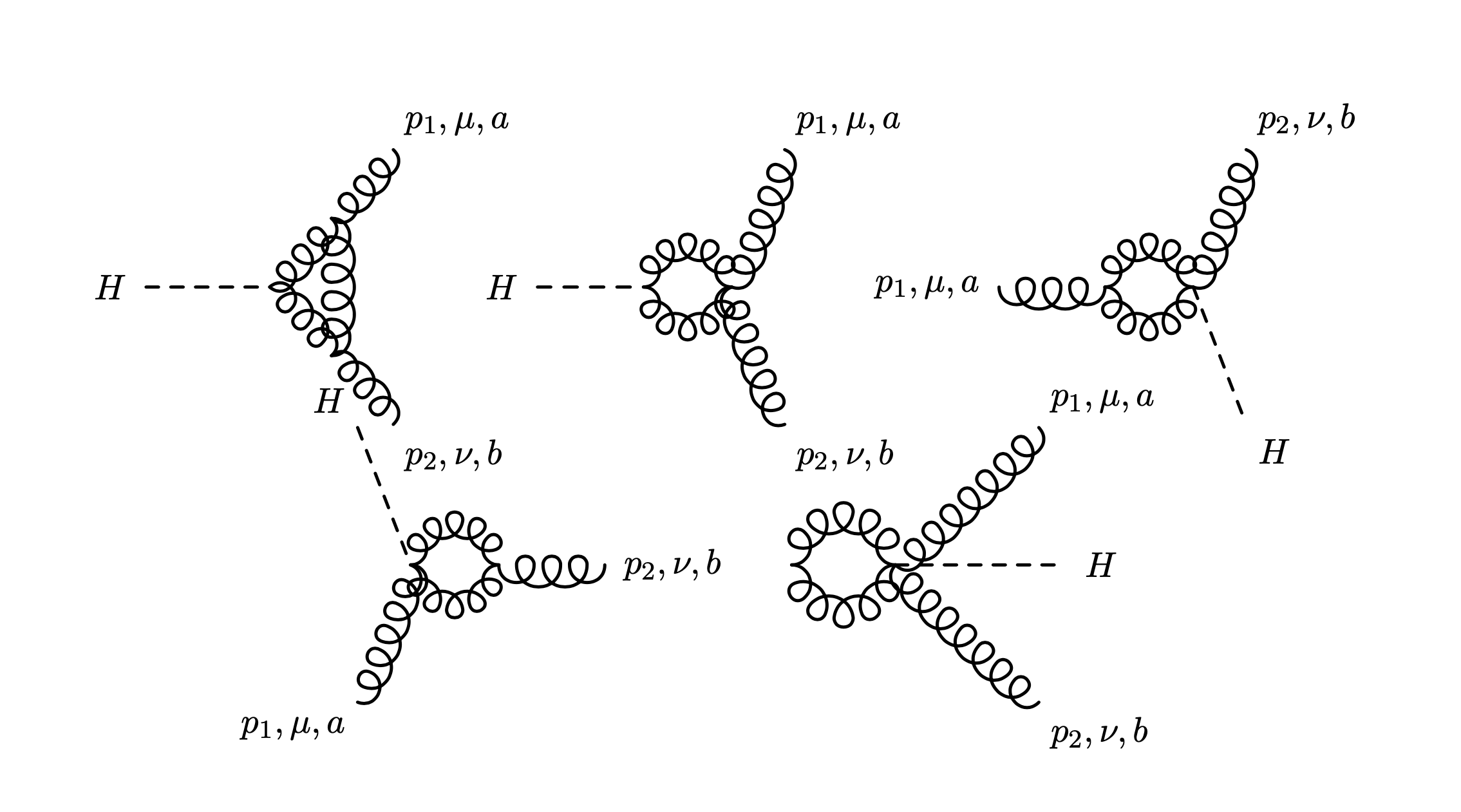}
    \caption{Virtual diagrams contribution to the decay rate $H \longrightarrow gg(g)$. From left to right they are respectively $V_1$,$V_2$,$V_3$,$V_4$,$V_5$. The dashed line represents the Higgs field, the curly lines represent the gluon field.}
\label{fig4.2}
\end{figure}

We can now compute the one-loop virtual diagrams. There are 5 diagrams that contribute to the one-loop order correction, which are represented in figure \ref{fig4.2}\footnote{The diagrams are depicted using \textit{FeynArts} \citep{2001CoPhC.140..418H}}. 

For all the diagrams we choose the external momenta of the two gluons to be $p_1$ and $p_2$ and the momentum of the Higgs boson to be \textit{q} and the internal momentum of the loop to be \textit{k}. All the external momenta are inwards, therefore we can write the equation of momentum-energy conservation as $p_1+p_2+q=0$. We apply the on-shell conditions by imposing $p_1^2=p_2^2=0$. The results for the integrals evaluated in this section can be found in Appendix \ref{ap:int}.

We begin with the diagram $V_1$ whose amplitude is given by
\begin{equation} \label{eq5.6}
\begin{split}
        V_1 & =- A g^2 C_A \delta^{ab} \\
        &\Bigg(  I_4 g^{\mu \nu} + 2I^{\alpha_1 \alpha_2} g^{\mu \nu} (p_1^{\alpha_1}p_1^{\alpha_2}  + p_2^{\alpha_1}p_2^{\alpha_2}) + 11 I_2^{\mu \nu} - I_2^\nu( 6p_1^\mu + p_2^\mu) + I_2^\mu (p_1 + 6 p_2)^\nu\\
         & - I (p_1 \cdot p_2) (4 p_2^\mu p_1^\nu + p_1^\mu p_2^\nu - 4(p_1 \cdot p_2)g^{\mu \nu}) + I_2 (-9 (p_1 \cdot p_2) g^{\mu \nu} + 9 p_2^\mu p_1^\nu - 3 p_1^\mu p_2^\nu + p_2^\mu p_2^\nu) \\
         & + 10 I^{\alpha_1 \mu \nu} (p_1 + p_2)^{\alpha_1} + I^{\alpha_1 \mu} (-2 p_1^{\alpha_1} p_1^\nu - 6 p_1^{\alpha_1} p_2^\nu-6 p_2^{\alpha_1} p_1^\nu + p_2^{\alpha_1} p_2^\nu)\\
         & + I^{\alpha_1 \nu} (-6 p_2^{\alpha_1} p_1^\mu - 2 p_2^{\alpha_1} p_2^\mu + p_1^{\alpha_1} p_1^\mu-6 p_1^{\alpha_1} p_2^\mu) \\
         & + I^{\alpha_1} (-6 p_1^{\alpha_1} p_2^\mu p_1^\nu + p_1^{\alpha_1} p_1^\mu p_2^\nu - 3 p_1^{\alpha_1} p_2^\mu p_2^\nu + 3 p_2^{\alpha_1} p_1^\mu p_1^\nu + 6 p_2^{\alpha_1} p_2^\mu p_1^\nu - p_2^{\alpha_1} p_1^\mu p_2^\nu\\
         & + 6 p_1 \cdot p_2 g^{\mu \nu} (p_1^{\alpha_1}-p_2^{\alpha_1})) + 3 I_2^{\alpha_1} g^{\mu \nu} (-p_1^{\alpha_1}  + p_2^{\alpha_1}) -(p_1 \cdot p_2) p_1^\mu I^{\nu} + I^{\mu} (p_1 \cdot p_2) p_2^\nu \Bigg)
\end{split}   
\end{equation}

We have already applied the conditions on-shell, $p_i^2=0$, $i=1,2$, and we adopted the following convention for the integrals

\begin{align}\label{eq:I}
    I^{\alpha_{1}\cdots\alpha_{n}} &= \int_{k}\frac{k^{\alpha_{1}}\cdots k^{\alpha_{n}}}{k^2(k-p_1)^2(k+p_2)^2}\\\label{eq:I2}
    I_{2}^{\alpha_{1}\cdots\alpha_{n}} &= \int_{k}\frac{k^{2} k^{\alpha_{1}}\cdots k^{\alpha_{n}}}{k^2(k-p_1)^2(k+p_2)^2}\\\label{eq:I4}
    I_{4}^{\alpha_{1}\cdots\alpha_{n}} &= \int_{k}\frac{k^{4} k^{\alpha_{1}}\cdots k^{\alpha_{n}}}{k^2(k-p_1)^2(k+p_2)^2}
\end{align}

After using the results collected in Appendix \ref{ap:int}, the final result reads

\begin{equation}
\begin{split}
    V_1 & = A g^2 C_A \delta^{ab} \Big[(-I_{quad}(\mu^2) (\dfrac{13}{2} g^{\mu \nu}) \\
& -I_{log}(\mu^2) (- \dfrac{43}{6} p_1 \cdot p_2 g^{\mu \nu} + \dfrac{1}{4} p_1^\mu p_1^\nu - \dfrac{1}{6} p_1^\mu p_2^\nu + \dfrac{29}{6}p_1^\nu p_2^\mu + \dfrac{1}{4}p_2^\mu p_2^\nu))\\
& + \dfrac{1}{12} \ln(-\mu_0)(-2 p_1^\nu p_2^\mu - \dfrac{13}{2} p_1 \cdot p_2) + \ln(-\mu_0)^2 (- p_1^\nu p_2^\mu + p_1 \cdot p_2)\\
&- \dfrac{5}{18} ( p_1^\nu p_2^\mu + 4 p_1 \cdot p_2)\Big]
\end{split}
\end{equation}
where $\mu_{0}=\mu^{2}/m_{H}^{2}$.

For the diagram $V_2$ we obtain

\begin{equation} \label{eq5.11}
\begin{split}
V_2 =& A g^2 C_A \delta^{ab} \dfrac{1}{2} \int_k \dfrac{1}{k^2(k-p_1-p_2)^2} \\
&\Big( 4 k^2 g^{\mu \nu} - 4k \cdot (p_1+p_2) g^{\mu \nu} + 2k^\mu k^\nu -k^\mu (p_1^\nu + p_2^\nu) -k^\nu (p_1^\mu + p_2^\nu)
\Big)
\end{split}
\end{equation}
and the regularized amplitude is

\begin{equation}
\begin{split}
V_2 =& A g^2 C_A \delta^{ab} \Big[ I_{quad}(\mu^2) (\dfrac{5}{2} g^{\mu \nu}) \\
& + \dfrac{1}{2} I_{log}(\mu^2) (- \dfrac{13}{3} p_1 \cdot p_2 g^{\mu \nu} - \dfrac{1}{3} p_1^\mu p_1^\nu - \dfrac{1}{3}p_1^\mu p_2^\nu - \dfrac{1}{3}p_1^\nu p_2^\mu - \dfrac{1}{3}p_2^\mu p_2^\nu)\\
& + \dfrac{1}{12} \ln(-\mu_0)(2 p_1^\nu p_2^\mu - \dfrac{13}{2} p_1 \cdot p_2) + \dfrac{5}{18} ( p_1^\nu p_2^\mu + 4 p_1 \cdot p_2)\Big]
\end{split}
\end{equation}

For the diagram $V_3$ we obtain

\begin{equation} 
\begin{split}
V_3 = & \dfrac{1}{2} Ag^2C_A \delta^{ab} \int_k \dfrac{1}{k^2(k-p_2)^2}\Big(2k^2 g^{\mu \nu} -2 k \cdot p_2 g^{\mu \nu} -3 p_1 \cdot p_2 g^{\mu \nu} \\
 & + 2 p_2^2 g^{\mu \nu} + 10 k^\mu k^\nu -5 k^\nu p_{2}^\mu  -5 k^\mu p_{2}^\nu + 3 p_{1}^\nu p_{2}^\mu + p_{2}^\mu p_{2}^\nu \Big)
\end{split}
\end{equation}
and regularizing
\begin{equation}
\begin{split}
V_3 =& A g^2 C_A \delta^{ab}\Big[ (I_{quad}(\mu^2) (\dfrac{7}{2} g^{\mu \nu}) \\
& + \dfrac{1}{2} I_{log}(\mu^2) (3 p_1^\nu p_2^\mu - 3p_1 \cdot p_2 g^{\mu \nu} - \dfrac{2}{3} p_2^\mu p_2^\nu)\Big]
\end{split}
\end{equation}

The diagram $V_4$ can be obtained from the result of $V_3$ by applying the replacements $p_{1}\leftrightarrow p_{2}$, $\mu \leftrightarrow \nu$

\begin{equation}
\begin{split}
V_4 =& A g^2 C_A \delta^{ab}\Big[( I_{quad}(\mu^2) (\dfrac{7}{2} g^{\mu \nu}) \\
& + \dfrac{1}{2} I_{log}(\mu^2) (3 p_1^\nu p_2^\mu - 3p_1 \cdot p_2 g^{\mu \nu} - \dfrac{2}{3} p_1^\mu p_1^\nu)\Big]
\end{split}
\end{equation}

Finally, the amplitude of diagram $V_5$ is given by
\begin{equation}
    V_{5}=A g^2 C_A \delta^{ab} \int_k \dfrac{-3 g^{\mu \nu}}{k^2-\mu^2} = -3 A g^2 C_A \delta^{ab} I_{quad}(\mu^2) g^{\mu \nu}.
\end{equation}

Once all amplitudes are regularized, we obtain an UV divergent part given by
\begin{equation}
    V_{div} =  \dfrac{\alpha_s}{\pi} C_A\left[\dfrac{I_{log}(\mu^2)}{2b}\right]V_0\;.
\end{equation}
It is worth noticing that quadratic divergent integrals cancel in the sum, being a proof of consistency of the method. Finally, the UV finite part amounts to (as can be easily checked, only the diagrams $V_1$ and $V_2$ contribute)
\begin{equation} \label{V_fin}
    V_{rest} =  \dfrac{\alpha_s}{\pi}C_{A}  \left[-\dfrac{\ln(\mu_0)^2}{4} -  \dfrac{i \pi \ln(\mu_0)}{2} + \dfrac{\pi^2}{4}\right]V_0.
\end{equation}
where we used $\ln(-\mu_0)^2 = \ln(\mu_0)^2 + 2 i \pi \ln(\mu_0) - \pi^2$.

At this point we can check if our result is UV finite after adding the countertem obtained in the last subsection (eq. (\ref{count}))

\begin{equation}
    V_{ren} = V_{div}+V_{count} = \dfrac{\alpha_s}{b \pi} \left[ \Big( I_{log}(\lambda^2) - I_{log}(\mu^2) \Big) \Big( \dfrac{11}{12}C_A - \dfrac{1}{3} T_f N_F \Big) \right] V_0.
\end{equation}

Using the scale relation, eq. \ref{2.19}, we obtain
\begin{equation} \label{V_ren}
    V_{ren} = \dfrac{\alpha_s}{\pi} \left[ \Big( \dfrac{11}{12}C_A - \dfrac{1}{3} T_f N_F \Big) \ln\Big( \dfrac{\lambda^2}{\mu^2} \Big)  \right]V_0 ,
\end{equation}
rendering an UV finite result as expected. 

Finally, the virtual decay rate can be obtained by considering the sum of the tree-level amplitude with the one-loop radiative correction
\begin{equation}
    V = V_0 + V_{ren} + V_{rest}.
\end{equation}
By squaring this result up to the order $\alpha_{s}$  we obtain

\begin{equation}
    |V|^2 = |V_0|^2 \left[ 1 + \dfrac{\alpha_s}{\pi}\left( \Big(\dfrac{11}{6}C_A - \dfrac{2}{3} T_f N_F\Big)\ln \Big(\dfrac{\lambda^2}{\mu^2} \Big) +\dfrac{C_A}{2}\Big(-\ln(\mu_0)^2+ \pi^2\Big) \right)  \right] 
\end{equation}
and the virtual decay rate is given by
\begin{equation} \label{virtual}
    \Gamma_v = \Gamma_0 \left[ 1 + \dfrac{\alpha_s}{\pi}\left( -\Big(\dfrac{11}{6}C_A - \dfrac{1}{3} N_F\Big)\ln(\mu_0) +\dfrac{C_A}{2}\Big(-\ln(\mu_0)^2+ \pi^2\Big) \right)  \right] .
\end{equation}
where we used $T_{F}=1/2$, and we are choosing the renormalization scale at the Higgs mass ($\lambda^{2}=m_{H}^{2}$) since $\mu_{0}=\mu^{2}/m_{H}^{2}$.

\subsection{Real Decay Rate} \label{Real Decay Rate}

\begin{figure}[h!]
    \centering
    \includegraphics[width=\textwidth]{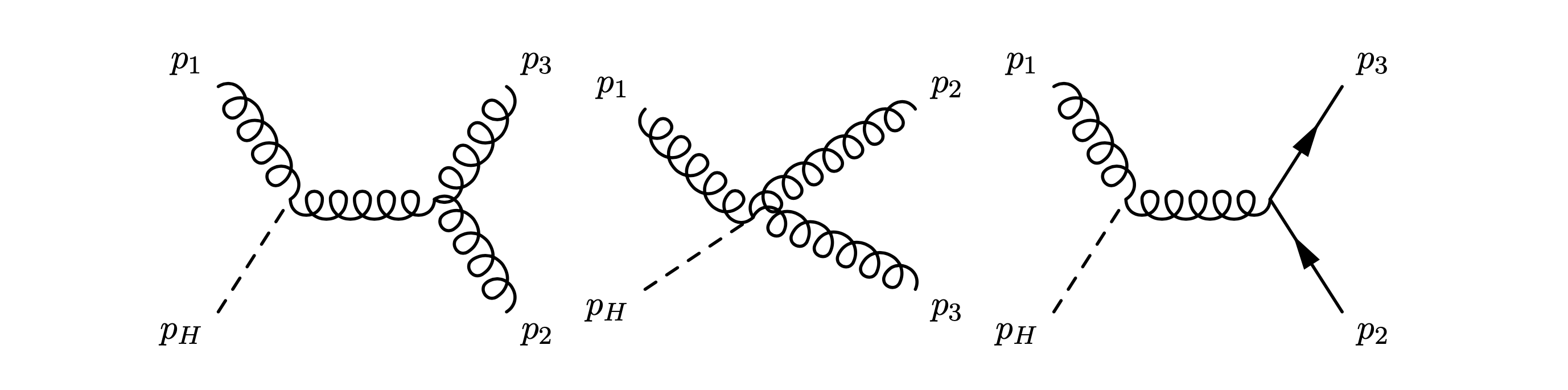}
\caption{Real diagrams contributing to the decay $H \longrightarrow ggg$ and $H \longrightarrow gq\bar{q}$. From left to right they are respectively $R_1$, $R_2$ and $R_3$. The dashed line represents the Higgs field and the curly lines represent the gluon field. The $\{p_i,p_j,p_k\}$ correspond to the three permutations of $p_i$, $p_j$ and $p_k$, so $R_1$ stands for 3 diagrams.}
\label{fig4.3}
\end{figure}

The diagrams that will contribute for the real decay until the $\alpha_s$ order are represented in figure \ref{fig4.3}.

We consider first the diagrams with gluon as external legs only, which is more involved to be obtained. In order to simplify the calculation we will adopt the spinor helicity formalism in this case. For the diagram with external (light) quarks, we will use the standard procedure. We begin with the \textit{s-channel} diagram $R_1$, whose amplitude is given by

\begin{equation} 
\begin{split}
iM_{R_1} & = g_{s}f^{bcd}V^{\nu \tau \delta}(p_2,p_3,-(p_2+p_3))  i \dfrac{g^{\delta \delta^{'}} \delta^{d d^{'}}}{(p_2+p_3)^2} iA \delta^{ad^{'}} H^{\delta^{'} \mu}(-(p_2+p_3),p_1)  \epsilon_1^{\mu} \epsilon_2^{\nu} \epsilon_3^{\tau}\\
& = -Ag_{s}f^{bca} \dfrac{1}{(p_2+p_3)^2} V^{\nu \tau \delta} H^{\delta \mu} \epsilon_1^{\mu} \epsilon_2^{\nu} \epsilon_3^{\tau}
\end{split}
\end{equation}
where the expanded tensors are given by
\begin{equation}
   V^{\nu \tau \delta}(p_2,p_3,-(p_2+p_3)) = (p_2-p_3)^\delta g^{\nu \tau} - p_2^\nu g^{\tau \delta} + p_3^\tau g^{\delta \nu}
\end{equation}
and
\begin{equation}
    H^{\delta \mu}(-(p_2+p_3),p_1) = - g^{\mu \delta}p_1 \cdot (p_2+p_3) + p_1^\delta (p_2+p_3)^\mu .
\end{equation}

Contracting the indices and using the Lorenz condition $\epsilon^\mu p_\mu=0$ we obtain
\begin{equation}
\begin{split}
iM_{R_1} = & \dfrac{-Ag_{s}f^{bca}}{s_{23}} (s_{12}+s_{13})(\epsilon_1 \cdot \epsilon_3 \: p_3 \cdot \epsilon_2 - p_2 \cdot \epsilon_3 \: \epsilon_1 \cdot \epsilon_2 )\\
& - s_{12} \: p_3 \cdot \epsilon_1 \: \epsilon_2 \cdot \epsilon_3 + s_{13} \: p_2 \cdot \epsilon_1 \: \epsilon_2 \cdot \epsilon_3\\
& + 2 (p_2 \cdot \epsilon_1 + p_3 \cdot \epsilon_1)  (p_1 \cdot \epsilon_2 \: p_2 \cdot \epsilon_3 - p_1 \cdot \epsilon_3 \: p_3 \cdot \epsilon_2).
\end{split}
\end{equation}

We apply now the spinor helicity formalism. Firstly we define three auxiliary momenta $r_i$, $i=1,2,3$, one for each of the massless gluon
\begin{align} \label{eq6.8}
r(\epsilon_1)&=p_3 \equiv 3, & r(\epsilon_2)&=p_1 \equiv 1, & r(\epsilon_3)&=p_2 \equiv 2.
\end{align}
By doing this choice the terms $p_2 \cdot \epsilon_3=p_1 \cdot \epsilon_2=p_3 \cdot \epsilon_1=0$ are automatically zero, which allows for a great deal of simplification of the amplitude,

\begin{equation}
\begin{split}
iM_{R_1} =& \dfrac{-Ag_{s}f^{bca}}{s_{23}}\\
& \Big[ (s_{12}+s_{13})(\epsilon_1 \cdot \epsilon_3 \: p_3 \cdot \epsilon_2) + s_{13} \: p_2 \cdot \epsilon_1 \: \epsilon_2 \cdot \epsilon_3 - 2 p_2 \cdot \epsilon_1 \: p_1 \cdot \epsilon_3 \: p_3 \cdot \epsilon_2 \Big].
\end{split}
\end{equation}
The \textit{t} channel is obtained by making the replacements $1 \longleftrightarrow 3$, $2 \longleftrightarrow 1$ and $3 \longleftrightarrow 2$ while the \textit{u} channel is obtained by making $1 \longleftrightarrow 2$, $2 \longleftrightarrow 3$ and $3 \longleftrightarrow 1$.

The amplitude for $R_2$ can be obtained in a similar fashion. The end result is
\begin{equation}
iM_{R_2} = -Ag_{s}f^{abc} \Big(p_1 \cdot \epsilon_3\: \epsilon_1 \cdot \epsilon_2 + p_2 \cdot \epsilon_1\: \epsilon_2 \cdot \epsilon_3 + p_3 \cdot \epsilon_2\: \epsilon_1 \cdot \epsilon_3 \Big).
\end{equation}

Summing the \textit{s}, \textit{t} and \textit{u} channels from $R_1$ with the one vertex diagram from $R_2$ we have

\begin{equation} 
\begin{split}
iM= - Ag_s \Bigg( & \dfrac{f^{bca}}{s_{23}} \big[(s_{12}+s_{13})(\epsilon_1 \cdot \epsilon_3 \: p_3 \cdot \epsilon_2 ) + s_{13} \: p_2 \cdot \epsilon_1 \: \epsilon_2 \cdot \epsilon_3 - 2 p_2 \cdot \epsilon_1 \: p_1 \cdot \epsilon_3 \: p_3 \cdot \epsilon_2 \big] \\
+ & \dfrac{f^{abc}}{s_{12}} \big[(s_{13}+s_{23}) (\epsilon_3 \cdot \epsilon_2 \: p_2 \cdot \epsilon_1 ) + s_{23} \: p_1 \cdot \epsilon_3 \: \epsilon_1 \cdot \epsilon_2 - 2 p_1 \cdot \epsilon_3 \: p_3 \cdot \epsilon_2 \: p_2 \cdot \epsilon_1 \big]\\
+ & \dfrac{f^{cab}}{s_{13}} \big[ (s_{23}+s_{12})(\epsilon_1 \cdot \epsilon_2 \: p_1 \cdot \epsilon_3) + s_{12} \: p_3 \cdot \epsilon_2 \: \epsilon_3 \cdot \epsilon_1 - 2 p_3 \cdot \epsilon_2 \: p_2 \cdot \epsilon_1 \: p_1 \cdot \epsilon_3 \big]\\
+ & f^{abc} ( p_1 \cdot \epsilon_3\: \epsilon_1 \cdot \epsilon_2 + p_2 \cdot \epsilon_1\: \epsilon_2 \cdot \epsilon_3 + p_3 \cdot \epsilon_2\: \epsilon_1 \cdot \epsilon_3 ) \Bigg).
\end{split}
\end{equation}

To obtain the unpolarized absolute square of the amplitude we need to sum over all possible colors and helicities,  
\begin{equation}
    |\overline{M}|^2= \sum_{col, polr}|M|^2 =\sum_{col} 2(|M^{+++}|^2+|M^{+--}|^2+|M^{-+-}|^2+|M^{--+}|^2)
\end{equation}

\noindent and use the spinor helicity formalism (see e.g. \cite{1997PhRvD..55.4005K}) to perform the spin sums. A massless real valued four momentum vector $p^\mu$ in the representation 

\begin{equation}
    p^{\alpha \dot{\alpha}}= \sigma_\mu^{\alpha \dot{\alpha}}p^\mu, \quad p_{ \dot{\alpha} \alpha} = \bar{\sigma}_{\dot{\alpha} \alpha}^\mu p_\mu
\end{equation}
where $\sigma_\mu = \big( \mathbb{I}, \Vec{\sigma} \big)$ and $\bar{\sigma}^\mu = \big( \mathbb{I},- \Vec{\sigma} \big)$ allows a bispinor decomposition, 

    \begin{equation}
        p^{\alpha \dot{\alpha}} = p \rangle [p, \quad  p_{ \dot{\alpha} \alpha} = p] \langle p
    \end{equation}
with the properties

\begin{equation}
    \langle p q \rangle = \sqrt{2 p \cdot q} e^{i \phi}, \quad  [ q p ] = \sqrt{2 p \cdot q} e^{-i \phi}
\end{equation}
for an arbitrary $\phi$ phase and

\begin{equation}
    p \cdot q = q^\mu p_\mu = \dfrac{1}{2} q_{\dot{\alpha} \alpha} p^{\alpha \dot{\alpha}} = \dfrac{1}{2} \langle q p \rangle [p q].
\end{equation}
The relation between these objects and the usual Mandelstam variables which we use to compute amplitudes is

\begin{equation} \label{eq3.20}
    \langle i j \rangle [j i] = 2p_i p_j = (p_i + p_j)^2 = s_{ij}
\end{equation}
The polarization vectors are given in this notation by 

\begin{equation}
    [\epsilon_p^-(r)]^{\alpha \dot{\alpha}} = \sqrt{2}  \dfrac{p \rangle [r}{[pr]},
\end{equation}

\begin{equation}
    [\epsilon_p^+(r)]^{\alpha \dot{\alpha}} = \sqrt{2}  \dfrac{r \rangle [p}{\langle r p \rangle}.
\end{equation}
from which follow the inner products

\begin{equation}
\begin{split}
    &\epsilon^-_{p}(r_1) \cdot \epsilon^+_q(r_2)= \dfrac{\langle p r_2 \rangle [q r_1]}{[p r_1] \langle r_2 q \rangle}\\
    & \epsilon^-_{p}(r_1) \cdot \epsilon^-_q(r_2) = \dfrac{\langle p q \rangle [r_2 r_1]}{[p r_1] [ 
    q r_2]}\\
    &\epsilon^+_{p}(r_1) \cdot \epsilon^+_q(r_2) = \dfrac{[ r_1 r_2 ] \langle q p \rangle}{\langle r_1 p \rangle \langle r_2 q \rangle}
\end{split}
\end{equation}

\begin{equation}
    \epsilon_p^-(r_1) \cdot q = \dfrac{1}{\sqrt{2}} \dfrac{\langle p q \rangle [q r_1]}{[p r_1]}, \quad \epsilon_p^+(r_1) \cdot q = \dfrac{1}{\sqrt{2}} \dfrac{\langle p q \rangle [q r_1]}{\langle r_1 p \rangle}
\end{equation}
with \textit{p}, \textit{q} being momenta and $r_1$ and $r_2$ reference momenta. 

Using our previous choice of reference momenta, we have for the ${+ - -}$ helicity
\begin{equation}
\begin{split}
    M^{+ - -} = & -Ag_{s}f^{abc} \Bigg[ \Big(\dfrac{\langle23\rangle}{[32]} \dfrac{\langle23\rangle[12]}{\langle31\rangle}\Big) \Big( \dfrac{s_{13}}{s_{23}} + \dfrac{s_{13}}{s_{12}} + \dfrac{s_{23}}{s_{12}} + 1\Big)\\
    & -2\Big(\dfrac{1}{\sqrt{2}} \dfrac{\langle23\rangle[12]}{\langle31\rangle}\Big)\Big(\dfrac{1}{\sqrt{2}} \dfrac{\langle13\rangle[21]}{[32]}\Big)\Big(\dfrac{1}{\sqrt{2}} \dfrac{\langle32\rangle[13]}{[21]}\Big)\Big(\dfrac{1}{s_{23}} + \dfrac{1}{s_{12}} + \dfrac{1}{s_{13}}\Big) \Bigg]
\end{split}
\end{equation}
and for the ${+ + +}$ helicity

\begin{equation}
\begin{split}
    M^{+ + +} & = -Ag_{s}f^{abc} \\
    & \Bigg[ \dfrac{[13]}{\langle31\rangle} \dfrac{1}{\sqrt{2}} \dfrac{\langle31\rangle[23]}{\langle12\rangle} \Big(\dfrac{s_{12}+s_{13}}{s_{23}} + \dfrac{s_{12}}{s_{13}} + 1 \Big)\\
    & + \dfrac{[23]}{\langle32\rangle} \dfrac{1}{\sqrt{2}} \dfrac{\langle23\rangle[12]}{\langle31\rangle} \Big(\dfrac{s_{13}+s_{23}}{s_{12}} + \dfrac{s_{13}}{s_{23}} + 1 \Big) \\
    & + \dfrac{[12]}{\langle21\rangle} \dfrac{1}{\sqrt{2}} \dfrac{\langle12\rangle[31]}{\langle23\rangle} \Big(\dfrac{s_{23}+s_{12}}{s_{13}} + \dfrac{s_{23}}{s_{12}} + 1 \Big)\\
    & + \dfrac{1}{\sqrt{2}} \dfrac{\langle23\rangle[12]}{\langle31\rangle} \dfrac{1}{\sqrt{2}} \dfrac{\langle12\rangle[31]}{\langle23\rangle} \dfrac{1}{\sqrt{2}} \dfrac{\langle31\rangle[23]}{\langle12\rangle} \Big(\dfrac{1}{s_{12}} + \dfrac{1}{s_{13}} + \dfrac{1}{s_{23}}\Big) \Bigg].
\end{split}
\end{equation}
Squaring the amplitudes, we obtain respectively

\begin{equation}
    |M^{+ - -}|^2 = \dfrac{1}{2} A^2g_{s}^2 f^2 \dfrac{s_{23}^3}{s_{12}s_{13}}, \quad 
    |M^{+ + +}|^2 = \dfrac{1}{2} A^2g_{s}^2 f^2 \dfrac{m_H^8}{s_{12} s_{13} s_{23}}.
\end{equation}
where we used the momentum-energy conservation $m_H^2 = s_{12} + s_{13} + s_{23}$.

The remaining helicity configurations can be obtained by permutation of momenta. The structure constants can be written as $f^2 = f_{abc} f^{abc} = 2 C_A^2 C_F$ with $C_F = \dfrac{N^2 - 1}{2N}$ and $C_A = N$ with $N=3$ for the $SU(3)$ group. We also have defined $\alpha_s = g_{s}^2/4 \pi$. Finally, we obtain for the unpolarized amplitude considering gluons only, $M_g$,

\begin{equation} 
    |\overline{M_g}|^2 = A^2 192 \pi \alpha_s \dfrac{1}{s_{12} s_{13} s_{23}} (s_{12}^4 + s_{13}^4 + s_{23}^4 + m_H^8).
\end{equation}

For the case of light quarks in the external legs, diagram $R_3$, the calculation is easier. Its amplitude $M_q$ is given by
\begin{equation}
    M_q = i A \alpha_s t_b  \epsilon_\nu(p_3) H^{\rho \nu}((-p_1-p_2),p_3) \Bar{u}^s(p_1) \gamma_\rho v^{s'}(p_2) \dfrac{1}{(-p_1-p_2)^2 + i \epsilon}. 
\end{equation}

By considering the unpolarized amplitude, one must sum over color and spin of the external particles. In this case, it is essential to recall that, in IReg, the IR divergences are parametrized as fictional masses for otherwise massless particles. Thus, when using the completeness relation for the sum over spin, one must retain the massive contribution. By doing so, one obtains
\begin{equation}\label{eq:quark}
    |\overline{M_q}|^2 = A^2 \alpha_s 4 \pi \text{Tr}(t_b t^b) \Big( \dfrac{(s_{13}^2 + s_{12}^2)}{s_{23}} + \dfrac{4 \mu^2}{s_{23}^2}  \dfrac{(s_{13} + s_{12})^2}{2} \Big)
\end{equation}
For the SU(3) group we have $\text{Tr}(t^b t_b) = \text{Tr} \Big( \dfrac{4}{3} \mathbb{I} \Big) = 4$.

The total unpolarized amplitude of the decay is finally given by

\begin{equation}
    |\overline{M}|^{2} = |\overline{M_g}|^{2} + |\overline{M_q}|^{2}
\end{equation}
Our next task is to perform the phase space integral considering the external particles (gluons and light quarks) with mass $\mu$. As customary, one defines dimensionless variables which, in our case, are given by
\begin{equation}
     s_{13} = q^2(\chi_1 + \mu_0),\quad s_{23} = q^2(\chi_2 + \mu_0),\quad s_{12} = q^2(1-\chi_1-\chi_2 + \mu_0),
\end{equation}
where  $\mu_0 = \dfrac{\mu^2}{q^2}$ and $q^{2}=m_{H}^{2}$ (we are using the frame of reference in which the Higgs boson is at rest). In terms of these variables the decay rate is given by

\begin{equation}
\begin{split}
    &\Gamma_r (H \longrightarrow gg(g), gq \Bar{q})  =\\
    & \Gamma_0  \dfrac{\alpha_s}{\pi} \int \Big[  3\Big(2 + 3 \chi_2 - \dfrac{4}{(\chi_2 + \mu_0)} + \dfrac{5\chi_1}{(\chi_2 + \mu_0)} - \dfrac{\chi_1^2}{(\chi_2 + \mu_0)} + \dfrac{1}{(\chi_1 + \mu_0) (\chi_2 + \mu_0)} \Big)\\ 
    & + N_F \Big( \dfrac{2\mu_0}{(\chi_2 + \mu_0)^2} + \dfrac{1}{(\chi_2 + \mu_0)} - \dfrac{2 \chi_1}{(\chi_2 + \mu_0)} + \dfrac{2\chi_1^2}{(\chi_2 + \mu_0)} -2 + 3\chi_2\Big) \Big] d\chi_1 d\chi_2
\end{split}
\end{equation}

The integration is over a massive phase space and we already used the energy momentum conservation condition $\chi_1+\chi_2+\chi_3=1$.
The integrals are evaluated in Appendix \ref{integrals} using  results collected in \cite{2014EPJC...74.2686P,2017EPJC...77..471G}. Notice we have already extracted the tree-level decay rate, which is given by
\begin{equation} \label{gamma_0}
    \Gamma_0 = \dfrac{|M_{Hgg}|^2}{32 \pi m_H} = \dfrac{A^2 m_H^3}{8 \pi}.
\end{equation}

The end result is
\begin{equation} \label{real}
\begin{split}
      & \Gamma_r (H \longrightarrow gg(g), gq \Bar{q}) = \\
      & \Gamma_0 \dfrac{\alpha_s}{\pi}\left[ 
    3 \Big(\dfrac{73}{12} + \dfrac{11}{6} \ln (\mu_0 ) + \dfrac{\ln^2(\mu_0 )}{2}  - \dfrac{\pi^2}{2}  \Big) + N_{F} \Big( \dfrac{-\ln{(\mu_0)}}{3} - \dfrac{7}{6} \Big)\right].  
\end{split}
\end{equation}

By combining equations \ref{virtual} and \ref{real}, we get the final decay rate for the process $H\longrightarrow gg(g)$

\begin{equation}
    \Gamma_T((H \longrightarrow gg(g), gq \Bar{q})) = \Gamma_v + \Gamma_r = \Gamma_0 \Big( 1 + \dfrac{\alpha_s}{\pi}  \Big( \dfrac{73}{4}  - \dfrac{7}{6}N_{F} \Big)\Big)
\end{equation}
We now take into account the correction of $A = \dfrac{\alpha_s}{3 \pi v}(1 + \dfrac{11}{4} \dfrac{\alpha_s}{\pi})$. We have that $\Gamma^0 \propto A^2$ and only keeping terms until order $\alpha_s^{3}$ our final result reads
\begin{equation}
    \Gamma_T((H \longrightarrow gg(g), gq \Bar{q})) = \Gamma_0 \left[ 1 + \dfrac{\alpha_s}{\pi}\left(\dfrac{95}{4}  - \dfrac{7}{6}N_{F} \right)\right],
\end{equation}
\noindent
complying with the well-known result of the literature\;\cite{1995NuPhB.453...17S}. As a final comment, in IReg it is easy to obtain the decay rate at any other desired renormalization point. This is achieved by leaving the renormalization scale $\lambda^2$ as a free parameter until the very end
\begin{equation}
    \Gamma_T((H \longrightarrow gg(g), gq \Bar{q})) = \Gamma_0 \left[ 1 + \dfrac{\alpha_s}{\pi}\left(\dfrac{95}{4}  - \dfrac{7}{6}N_{F} +\dfrac{33 - 2N_{F}}{6} \ln \left(\dfrac{\lambda^{2}}{m_{H}^{2}} \right)\right)\right].
\end{equation}

\section{Comparison with dimensional schemes}
\label{sec:comparison}

In this section we intend to compare our results with the ones obtained in the context of dimensional schemes, in a similar way to the discussions presented in 
%\cite{Gnendiger:2017pys, TorresBobadilla:2020ekr}
\cite{2017EPJC...77..471G, 2021EPJC...81..250T}. For the results in dimensional schemes, we will  mainly use the analysis performed in %\cite{Broggio:2015ata,Gnendiger:2014nxa}
\cite{Broggio2015ComputationO, 2014PhLB..733..296G}. In these references, the form factors for the decay $H\rightarrow gg$ were computed up to two-loop. As discussed there, in the case of dimensional schemes other than CDR it is necessary to take into account a broader set of operators in the effective Lagrangian. This is due to the presence of additional, fictitious particles, denoted $\epsilon$-scalars. Although the main differences arise when one performs the two-loop analysis, in DRED, the one-loop contribution due to light quarks can only be consistently obtained when additional operators are taken into account. As we have shown in the previous section, in the case of IReg the inclusion of $\epsilon$-scalars (or additional operators) is not necessary.

Following 
%\cite{Gnendiger:2014nxa,Broggio:2015ata}
\cite{2014PhLB..733..296G, Broggio2015ComputationO}, the form factors in all dimensional schemes already UV renormalized can be obtained. In CDR the form factor related to the process $H\rightarrow gg$ is given by
 \newcommand{\Neps}{N_\epsilon}
 \newcommand{\alphas}{\alpha_s}
 \def\leps{\lambda_{\epsilon}}
\def\gtilde{{\tilde{g}}}
\begin{equation}
F_{\scriptscriptstyle{\text{CDR}}}  (\alphas)
=\Big(\frac{\alphas}{4\pi}\Big)
\Bigg\{C_A \Bigg[
  -\frac{2}{\epsilon ^2}
  -\frac{11}{3\epsilon}
  +\frac{\pi ^2}{6}
  \Bigg]
+\frac{2 N_{F}}{3 \epsilon }\Bigg\}+\mathcal{O}(\epsilon).
\end{equation}
In the case of FDH, one needs also to consider the $\epsilon$-scalars when performing the UV renormalization, which amounts to 
 \begin{align}
   F_{\scriptscriptstyle{\text{FDH}}}  (\alphas)
 &=F_{\scriptscriptstyle{\text{CDR}}}  (\alphas) + \Big(\frac{\alphas}{4\pi}\Big)
   \Neps C_A \Bigg[
      1+\frac{1}{6\epsilon }
      \Bigg]
    +\mathcal{O}(\Neps\epsilon).
\end{align}
Finally, in DRED one must consider both processes $H\rightarrow gg$ and $H\to\gtilde\gtilde$ where $\gtilde$ stands for the $\epsilon$-scalars. In this case, one needs the form factor of FDH and also the form factor with $\epsilon$-scalars
\begin{equation}
F_{\gtilde\gtilde}  (\alphas)
=\Big(\frac{\alphas}{4\pi}\Big)
\Bigg\{C_A \Bigg[
  -\frac{2}{\epsilon ^2}
  -\frac{4}{\epsilon}
  +2+\frac{\pi ^2}{6}-2\Neps
  \Bigg]
+\frac{N_{F}}{3 \epsilon }\Bigg\}+\mathcal{O}(\Neps\epsilon).
\end{equation}
One should notice that we are already setting the couplings related to $\epsilon$-scalars equal to their corresponding values in usual QCD. This is justified because the UV renormalization was already carried out. In all cases, the form factors are normalized to their tree-level values. Recovering them, the virtual contribution to the decay rate can be readily obtained for each scheme after setting $\Neps=2\epsilon$
\begin{align}
    \Gamma_{v}^{\scriptscriptstyle{\text{CDR}}}=&\Gamma_{0}\left\{1+\frac{\alphas}{\pi}\left[C_{A}\left(-\frac{1}{\epsilon^{2}}-\frac{11}{6\epsilon}+\frac{\pi^{2}}{12}\right)+\frac{N_{F}}{3\epsilon}\right]\right\}+\mathcal{O}(\epsilon)
    \\
    \Gamma_{v}^{\scriptscriptstyle{\text{FDH}}}=&\Gamma_{0}\left\{1+\frac{\alphas}{\pi}\left[C_{A}\left(-\frac{1}{\epsilon^{2}}-\frac{11}{6\epsilon}+\frac{\pi^{2}}{12}+\frac{1}{6}\right)+\frac{N_{F}}{3\epsilon}\right]\right\}+\mathcal{O}(\epsilon)
    \\
     \Gamma_{v}^{\scriptscriptstyle{\text{DRED}}}=&\Gamma_{0}\left\{1+\frac{\alphas}{\pi}\left[C_{A}\left(-\frac{1}{\epsilon^{2}}-\frac{11}{6\epsilon}+\frac{\pi^{2}}{12}\right)+\frac{N_{F}}{3\epsilon}+\frac{N_{F}}{6}\right]\right\}+\mathcal{O}(\epsilon)
\end{align}

We can compare these results to eq.\ref{virtual}, and see that the correspondence $\epsilon^{-1}\rightarrow\log{\mu_{0}}$, $\epsilon^{-2}\rightarrow\log^{2}{\mu_{0}}/2$ is fulfilled as first noticed in 
%\cite{Gnendiger:2017pys}
\cite{2017EPJC...77..471G}. Moreover, the result in CDR does not have any finite term (apart from factors of $\pi^2$ that will be cancelled against the real contribution). The same statement holds true for IReg. For FDH and DRED, on the other hand, there is the appearance of finite terms proportional to $C_{A}$ and $N_{F}$ respectively. 

Regarding the real contributions, to the best of our knowledge, the results for all schemes is not available in the literature. Nevertheless, the part proportional to $N_{F}$ can be readily obtained. For CDR, there is only one diagram (the diagram on the right of fig. \ref{fig4.3}) which produces the following unpolarized amplitude
\begin{equation}
    |\bar{M}_{q}^{\scriptscriptstyle{\text{CDR}}}|^{2}\propto  \dfrac{(s_{13}^2 + s_{12}^2)}{s_{23}} + \dfrac{(d-4)}{2}\dfrac{(s_{13} + s_{12})^2}{s_{23}}.
\end{equation}

For FDH, we have the same diagram of CDR and obtain the same result. This happens because the diagram is not 1PI, implying that the internal gluon is regular in the notation of 
%\cite{Gnendiger:2017pys}
\cite{2017EPJC...77..471G}. Therefore, although the external gluon is split into a $d$-dimensional gluon and a $\epsilon$-scalar, there is no way to produce a diagram with only $\epsilon$-scalars. In the case of DRED this happens, since all vector bosons must be split. The unpolarized amplitude in DRED is given by
\begin{equation}
    |\bar{M}_{q}^{\scriptscriptstyle{\text{DRED}}}|^{2}\propto  \dfrac{(s_{13}^2 + s_{12}^2)}{s_{23}} + \dfrac{(d-4)}{2}\dfrac{(s_{13} + s_{12})^2}{s_{23}}+\dfrac{\Neps}{2}\dfrac{(s_{13} + s_{12})^2}{s_{23}}.
\end{equation}
It should be noticed that, by setting $\Neps=2\epsilon$ one obtains the result of a strictly four-dimensional calculation. One can also compare these results with eq. \ref{eq:quark} obtained within IReg which we reproduce below 
\begin{equation}
    |\bar{M}_{q}^{\scriptscriptstyle{\text{IReg}}}|^2 \propto  \dfrac{(s_{13}^2 + s_{12}^2)}{s_{23}} + 2 s \mu_{0}\dfrac{(s_{13} + s_{12})^2}{s_{23}^{2}}
\end{equation}
%For a better comparison, we have already removed terms that will not contribute to the decay rate in the limit $\mu_0\rightarrow0$. 
As can be seen, the result in IReg is similar to CDR/FDH in the sense that there is an extra term. In the case of CDR/FDH it comes from extending the physical dimension to $d$, while in IReg it is encoded in the fictitious mass that we have added for the massless particles. Once the unpolarized amplitude is known in all schemes, one can obtain the part proportional to $N_{F}$ of the real contribution to the decay rate 
\begin{align}
    \Gamma_{r}^{\scriptscriptstyle{\text{CDR/FDH}}}=&\Gamma_{0}\frac{\alphas}{\pi}\left[-\frac{1}{3\epsilon}-\frac{7}{6}\right]N_{F}+\mathcal{O}(\epsilon)
    \\
    \Gamma_{r}^{\scriptscriptstyle{\text{DRED}}}=&\Gamma_{0}\frac{\alphas}{\pi}\left[-\frac{1}{3\epsilon}-\frac{4}{3}\right]N_{F}+\mathcal{O}(\epsilon)
\end{align}
Once again, the result in IReg can be mapped to the one of CDR/FDH after the identification $\epsilon^{-1}\rightarrow\log{\mu_{0}}$.

\section{Conclusion} 
\label{sec:conclusion}

In conclusion, the decay rate  $ \Gamma((H \longrightarrow gg(g), gq \Bar{q}))$  at $\alpha_s^3$ order in the strong coupling and large top quark mass limit has been computed  using an effective interaction Lagrangian of Higgs to two gluons added to the QCD Lagrangian  in the framework of the fully quadri-dimensional regularization scheme IReg and compared to dimensional schemes CDR, FDH/DRED. The purpose has been two-fold. Firstly to achieve not only a full separation of BDI from the UV finite integrals (which IReg accomplishes to arbitrary loop order) but to single out the IR content as well, rendering the evaluation of the decay rate more amenable to numerical calculations. Secondly to verify whether the additional degrees of freedom associated to epsilon scalars in some of the dimensional schemes have a counterpart in the non-dimensional scheme IReg. 

The present calculation provided a proof of concept example involving a non-abelian effective theory Lagrangian with sufficient complexity to allow to infer that crucial steps of the procedure are in compliance with fundamental requirements such as gauge invariance and the removal of IR singularities fulfilling the KLN theorem. In particular the use of a mass regulator in the propagators is adequately implemented in IReg, as well as the renormalization schemes adopted for an effective theory. The latter point in particular is non-trivial, as it involves the use of the method’s renormalization scale relations impacting on the overall cancellation of IR singularities. In addition, by comparing with different dimensional schemes, one concludes that IReg does not require the use of evanescent fields at one loop level.

%
% APPENDIX
%
\appendix

\newpage
\section{Feynman rules for the effective Lagrangian} \label{Feyn_rules}
We collect the Feynman rules for the \textit{2-gluon-Higgs}, \textit{3-gluon-Higgs} and \textit{4-gluon-Higgs} with the effective interaction and the \textit{3-gluon}, \textit{4-gluon} interactions with the YM lagrangian. For convention all the external momenta are going inwards.

\subsection*{2-gluon-Higgs Feynman rule}
\begin{figure}[h!]
    \centering
    \includegraphics[width=\textwidth]{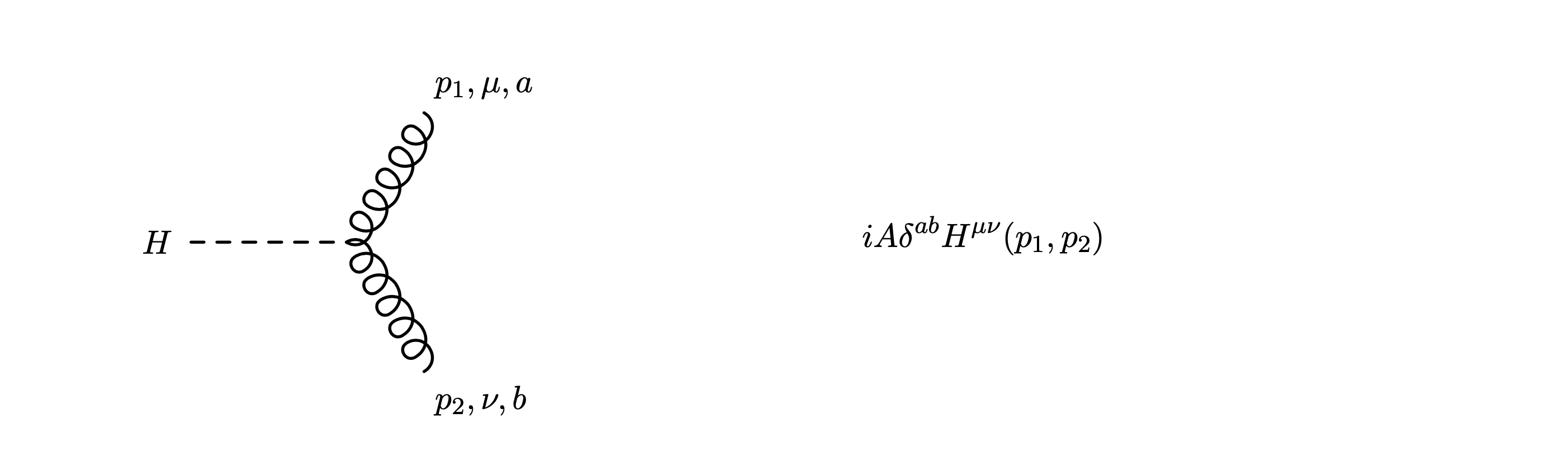}
\caption{\textit{2-gluon-Higgs} Feynman rule.} 
\label{2-gluon-Higgs Feynman rule}
\end{figure}

The \textit{2-gluon-Higgs} Feynman rule is represented in figure \ref{2-gluon-Higgs Feynman rule}.

\subsection*{3-gluon-Higgs Feynman rule}
\begin{figure}[h!]
    \centering
    \includegraphics[width=\textwidth]{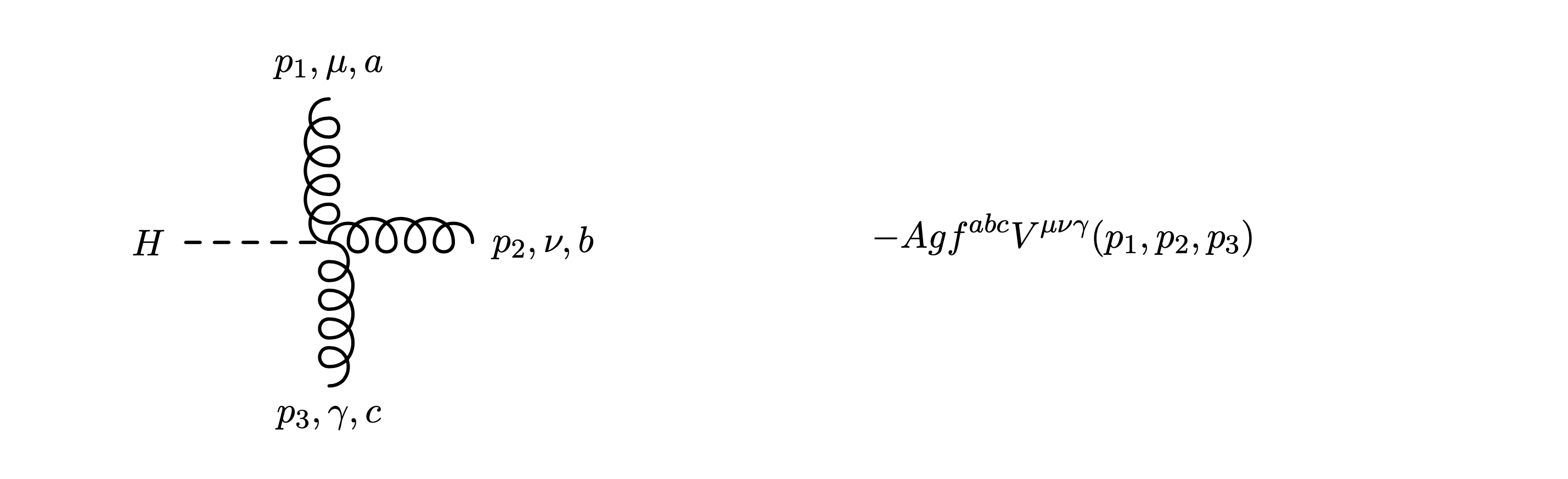}
\caption{\textit{3-gluon-Higgs} Feynman rule.} 
\label{3-gluon-Higgs Feynman rule}
\end{figure}

The \textit{3-gluon-Higgs Feynman rule} is represented in figure \ref{3-gluon-Higgs Feynman rule}.

where the tensor is $V^{\mu \nu \gamma}(p_1,p_2,p_3)=(p_1-p_2)^\gamma g^{\mu \nu} + (p_2-p_3)^\mu g^{\nu \gamma} + (p_3-p_1)^\nu g^ {\mu \gamma}.$

\subsection*{4-gluon-Higgs Feynman rule}

\begin{figure}[h!]
    \centering
    \includegraphics[width=\textwidth]{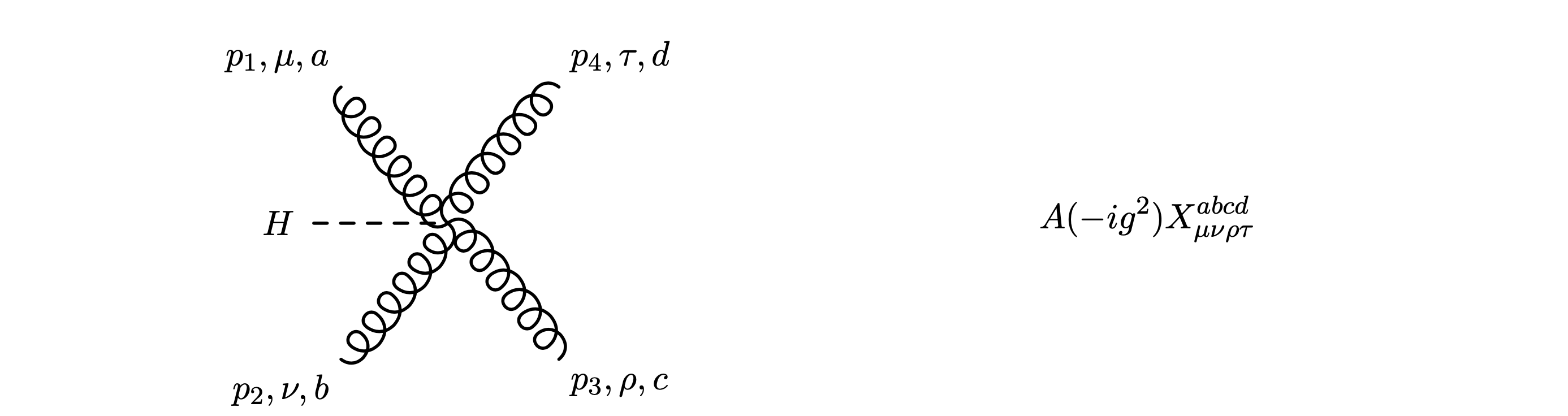}
\caption{\textit{4-gluon-Higgs} Feynman rule.} 
\label{4-gluon-Higgs Feynman rule}
\end{figure}

The \textit{4-gluon-Higgs Feynman rule} is represented in figure \ref{4-gluon-Higgs Feynman rule}.

with the tensor being

\begin{equation} \label{4.28}
\begin{split}
    X^{abcd}_{\mu \nu \rho \tau}= &f^{abe}f^{cde} (g_{\mu \rho}g_{\nu \tau}-g_{\mu \tau}g_{\nu \rho})\\
    +& f^{ace}f^{bde} (g_{\mu\nu}g_{\rho\tau}-g_{\mu\tau}g_{\rho\nu}) \\
    +&f^{ade}f^{bce} (g_{\mu\nu}g_{\tau\rho}-g_{\mu\rho}g_{\nu\tau}).
\end{split}
\end{equation}

For the \textit{3-gluon} and \textit{4-gluon} interactions, the Feynman rules coming from the Yang Mills Lagrangian are the same as of the effective vertices \textit{3-gluon-Higgs} and \textit{4-gluon-Higgs} apart from the effective coupling \textit{A}.

\section{Integrals used in the evaluation of the virtual contribution to \texorpdfstring{$H\rightarrow gg$}{TEXT}}
\label{ap:int}

In this appendix we collect the integrals needed for the evaluation of the virtual contribution to the process  $H\rightarrow gg$. We organize them in terms of the diagrams presented in figure \ref{fig4.2}. The UV finite part of the integrals were evaluated using \textit{Package-X} \citep{2015CoPhC.197..276P}.

%\subsection{Integrals of diagram $V_1$}
\subsection{Integrals of diagram \texorpdfstring{$V_1$}{TEXT}}

We define $b=\dfrac{i}{(4 \pi)^2}$ and apply the on-shell limit for each integral. We also omit all the integrals where we already applied the limit $\mu_0 = \mu^2/m_H^2 \rightarrow 0$.
The integrals are defined in equations \ref{eq:I} - \ref{eq:I4} and their results read

\begin{equation}
    I = b \dfrac{\ln^2{(-\mu_0)}}{4 p_1\cdot p_2}
\end{equation}

\begin{equation}
    I^{\mu} = - b \dfrac{2+ \ln{(-\mu_0)}}{(p_{1\mu} - p_{2\mu})}2 p_1 \cdot p_2
\end{equation}

\begin{equation}
\begin{split}
    I^{\mu \nu} = & \dfrac{I_{log}(\mu^2)}{4} g^{\mu \nu} + \dfrac{b}{4 p_1 \cdot p_2} \Big[ \ln (-\mu_0) p_1.p_2 g_{\mu \nu}+3 p_1.p_2 g_{\mu \nu }\\
    & - 2 p_{1\mu} \left((\ln (-\mu_0)+2) p_{1 \nu }+p_{2 \nu }\right)
    -2 \ln (-\mu_0) p_{2 \mu} p_{2 \nu}-2 p_{1 \nu } p_{2 \mu}-4 p_{2 \mu} p_{2 \nu} \Big]
\end{split}
\end{equation}

\begin{equation}
\begin{split}
    I^{\alpha \mu \nu} = & \dfrac{I_{log}(\mu^2)}{12} ((p_1-p_2)^\nu g^{\alpha \mu} + (p_1-p_2)^\mu g^{\alpha \nu} + (p_1 - p_2)^\alpha g^{\mu \nu} )\\
    & + b \Big[ \frac{1}{18 p_1 \cdot p_2} 4 (-p_1.p_2 p_{2 \mu} g_{\alpha \nu }+ p_{1 \nu } p_1 \cdot p_2 g_{\alpha \mu }-p_1 \cdot p_2 p_{2 \nu } g_{\alpha \mu }-p_1 \cdot p_2 p_{2 \alpha} g_{\mu \nu})\\
    & 3 \ln (-\mu_0)(- p_1 \cdot p_2 p_{2 \mu} g_{\alpha \nu }+ p_{1 \nu} p_1 \cdot p_2 g_{\alpha \mu }- p_1 \cdot p_2 p_{2 \nu } g_{\alpha \mu }) \\
    & + p_{1 \mu} ((3 \ln (-\mu_0)+4) p_1 \cdot p_2 g_{\alpha \nu }-3 p_{1 \nu } p_{2 \alpha}+3 p_{2 \alpha} p_{2 \nu })\\
    & -3 \ln (-\mu_0) p_1 \cdot p_2 p_{2 \alpha} g_{\mu \nu}+p_{1 \alpha} (4 p_1 \cdot p_2 g_{\mu \nu }+3 \ln (-\mu_0) p_1 \cdot p_2 g_{\mu \nu }-3 p_{1 \nu } p_{2 \mu}\\
    & -p_{1 \mu} ((6 \ln (-\mu_0)+13) p_{1 \nu}+3 p_{2 \nu})+3 p_{2 \mu} p_{2 \nu})+3 p_{1 \nu} p_{2 \alpha} p_{2 \mu }\\
    & +13 p_{2 \alpha} p_{2 \mu} p_{2 \nu}+6 \ln (-\mu_0) p_{2 \alpha} p_{2 \mu } p_{2 \nu}) \Big]
\end{split}
\end{equation}

\begin{equation}
    I_2 = I_{log}(\mu^2) + b( 2 + \ln{(-\mu_0)} )
\end{equation}

\begin{equation}
    I_2^{\nu} = \dfrac{I_{log}(\mu^2)}{2}(p_1^\nu - p_2^\nu) + \dfrac{b}{2}(2 +\ln{(-\mu_0)})(p_{1 \nu} - p_{2 \nu})
\end{equation}

\begin{equation}
\begin{split}
    I_2^{\mu \nu} = & \dfrac{I_{quad}(\mu^2)}{2} g^{\mu \nu} + \dfrac{I_{log}(\mu^2)}{3} (p_2^\mu p_2^\nu + p_1^\mu p_1^\nu) - \dfrac{I_{log}(\mu^2)}{6}(p_1 \cdot p_2 g^{\mu \nu} + p_1^\mu p_2^\nu + p_1^\nu p_2^\mu) \\
    & + \frac{b}{36} \Big[ \Big( \ln (-\mu_0) (12(p_{1 \mu} p_{1 \nu} + p_{2 \mu} p_{2 \nu})-6 (p_{1 \nu} p_{2 \mu} + p_{2 \mu} p_{2 \nu} + p_1 \cdot p_2 g_{\mu\nu }) \Big)\\
    & +26( p_{1 \mu} p_{1 \nu}+p_{2 \mu} p_{2 \nu})-10 (p_{1 \nu} p_{2 \mu}+p_{1 \mu} p_{2 \nu})- 16  p_1 \cdot p_2 g_{\mu \nu} ) \Big]
\end{split}
\end{equation}

\begin{equation}
    I_4 = I_{quad}(\mu^2) - I_{log}(\mu^2) p_1 \cdot p_2  - b (p_1 \cdot p_2) (2 + \ln{(-\mu_0)})
\end{equation}

\subsection{Integrals of diagram \texorpdfstring{$V_2$}{TEXT}}

\begin{equation}
\begin{split}
    \int_k \dfrac{1}{(k-p_1-p_2)^2} = \int_k \dfrac{1}{k^{2}-2k\cdot(p_1+p_2) + 2p_{1}\cdot p_{2}} = & I_{quad}(\mu^2) + 2b p_1 \cdot p_2 \ln(-\mu_0)
\end{split}
\end{equation}

\begin{equation}
\begin{split}
    \int_k \dfrac{k^\mu}{k^2 (k-p_1-p_2)^2} & = \int_k \dfrac{k^\mu}{k^{2}(k^{2}-2k\cdot(p_1+p_2) + 2p_{1}\cdot p_{2})} \\
    & = \dfrac{I_{log}(\mu^2)}{2} (p_1 + p_2)^\mu + \frac{b}{2} (\ln (-\mu_0)+2) \left(p_{1 \mu}+p_{2 \mu}\right)
\end{split}
\end{equation}

\begin{equation}
\begin{split}
    \int_k \dfrac{k^\mu k^\nu}{k^2(k-p_1-p_2)^2} =& \int_k \dfrac{k^\mu  k^\nu}{k^{2}(k^{2}-2k\cdot(p_1+p_2) + 2p_{1}\cdot p_{2})} \\
    =& \dfrac{I_{quad}(\mu^2)}{2} g^{\mu \nu} - \dfrac{I_{log}(\mu^2)}{2} p_1 \cdot p_2 g^{\mu \nu} \\
    & + \dfrac{I_{log}(\mu^2)}{3} (g^{\mu \nu} p_1 \cdot p_2 + (p_1 + p_2)^\mu (p_1 + p_2)^\nu)\\
    &+ b \Big[ \frac{1}{18} p_1\cdot p_2 g_{\mu \nu}+\frac{1}{3} \ln (-\mu_0)\big( p_1 \cdot p_2 g_{\mu \nu }+ p_{1 \mu } p_{1 \nu }\\
    & + p_{1 \nu} p_{2 \mu}+ p_{1 \mu} p_{2 \nu}+ p_{2 \mu } p_{2 \nu}\big)\\
    & +\frac{2}{9} \big(p_{1 \mu} p_{1 \nu} + p_{1 \nu} p_{2 \mu} + p_{1 \mu} p_{2 \nu} + p_{2 \mu} p_{2 \nu} \big) \Big]
\end{split}
\end{equation}

%\subsection{Integrals of diagrams $V_3$ and $V_4$}
\subsection{Integrals of diagrams \texorpdfstring{$V_3$}{TEXT} and \texorpdfstring{$V_4$}{TEXT}}

\begin{equation}
    \int_k \dfrac{1}{(k-p_2)^2} = \int_k \dfrac{1}{k^{2}-2k.p_2} = I_{quad}(\mu^2)
\end{equation}

\begin{equation}
    \int_k \dfrac{1}{k^2(k-p_2)^2} = \int_k \dfrac{1}{k^{2}(k^{2}-2k.p_2)} = I_{log}(\mu^2)
\end{equation}

\begin{equation}
    \int_k \dfrac{k^\mu}{k^2(k-p_2)^2} = \int_k \dfrac{k^\mu}{k^{2}(k^{2}-2k.p_2)} = \dfrac{I_{log}(\mu^2)}{2} p_2^\mu
\end{equation}

\begin{equation}
    \int_k \dfrac{k^\mu k^\nu}{k^2(k-p_2)^2} = \int_k \dfrac{k^\mu k^\nu}{k^{2}(k^{2}-2k.p_2)}=\dfrac{I_{quad}(\mu^2)}{2} g^{\mu \nu} + \dfrac{I_{log}(\mu^2)}{3} p_2^\mu p_2^\nu
\end{equation}

\section{Integrals used in the evaluation of the real contribution to \texorpdfstring{$H\rightarrow gg$}{TEXT}} \label{integrals}

The phase space integral is found to be

\begin{equation}
    \rho = \dfrac{q_0^2}{(4\pi)^3} \int_{\chi_1^{-}}^{\chi_1^{+}} d\chi_1 \int_{\chi_2^{-}}^{\chi_2^{+}} d\chi_2
\end{equation}
where the boundaries for $\chi_2^\pm$ are given by 
\begin{equation}
    \chi_2^\pm = \dfrac{1 - \chi_1}{2} \pm \sqrt{\dfrac{(\chi_1 - 3\mu_0)[(1-\chi_1)^2-4\mu_0]}{4(\chi_1 + \mu_0)}},
\end{equation}
and the boundaries for $\chi_1^\pm$ are
\begin{equation}
    \chi_1^{+} = 1 - 2\sqrt{\mu_0},\quad\mbox{and}\quad\chi_1^{-} = 3\mu_0.
\end{equation}

Finally, the integrals were evaluated using the following equations, \cite{2017EPJC...77..471G, 2014EPJC...74.2686P}  %Pittau, to d

\begin{equation}
    I(s) = \int d \chi_1 d \chi_2 \dfrac{1}{(\chi_1+\mu_0)(\chi_2+\mu_0)}
\end{equation}
and

\begin{equation}
    J_p(s) = \int d\chi_1 d\chi_2 \dfrac{\chi_1^p}{(\chi_2+\mu_0)}
\end{equation}
with $p\geq0$. Using our limit of integrations, the integrals are evaluated to be

\begin{equation} \label{6.60}
    I(s) = \dfrac{\ln^2(\mu_0)-\pi^2}{2}
\end{equation}
and

\begin{equation} \label{6.61}
\begin{split}
    J_p(s) =& - \dfrac{1}{p+1} \ln(\mu_0) + \int_0^1 d\chi_1 \chi_1^p [\ln(\chi_1) + 2 \ln(1-\chi_1)]\\
    =& - \dfrac{1}{p+1} \ln(\mu_0) - \dfrac{1}{p+1} \Big[\dfrac{1}{p+1} +2 \sum_{n=1}^{p+1} \dfrac{1}{n} \Big].
\end{split}
\end{equation}

%
% ACKNOWLEDGMENTS
%
\newpage
\acknowledgments
We acknowledge support from Fundação para a Ciência e Tecnologia (FCT) through the projects CERN \slash FIS-PAR \slash 0040 \slash 2019, CERN \slash FIS-COM \slash 0035 \slash 2019 and UID \slash FIS \slash 04564 \slash 2020 and National Council for Scientific and Technological Development – CNPq through projects 166523\slash 2020-8 and 302790 \slash 2020-9.

%-----------------------------------------------------------------
%
% BIBLIOGRAFIA
% 
%-----------------------------------------------------------------
%\bibliographystyle{apalike}
\bibliographystyle{unsrt}
\bibliography{main}
\end{document}